\documentstyle[aps,floats,epsfig]{revtex}
\begin{document}

\twocolumn
\renewcommand{\topfraction}{1.0}
\twocolumn[\hsize\textwidth\columnwidth\hsize\csname
@twocolumnfalse\endcsname
\title{Echoes of the fifth dimension?}
\author{Luis A. Anchordoqui, Thomas P. McCauley, Stephen Reucroft, and 
John Swain} 
\address{Department of Physics, Northeastern University, Boston, Massachusetts
02115}
\maketitle

\begin{abstract}
In this article we examine the question of whether the highest energy cosmic 
ray primaries 
could be ultra relativistic magnetic monopoles. The analysis is performed 
within the framework of large compact dimensions and TeV scale quantum 
gravity. Our study indicates that while this hypothesis must be regarded as 
highly speculative it cannot be ruled out with present data.

\end{abstract}

\vskip2pc]

This past year has seen a massive resurgence of interest in higher dimensional 
spacetimes \cite{renacimiento}, a key new concept being the localization of 
matter, and even gravity on branes embedded in extra 
dimensions \cite{trapping}. Depending 
on the dimensionality and the particular form of this space, the long standing
(Planck) hierarchy problem can find alternative solutions.
In the canonical example of \cite{nima}, the Planck scale of the four 
dimensional world is 
related to that of a higher dimensional space-time simply by a volume factor,
\begin{equation}
r = \left( \frac{M_{\rm pl}}{M_*} \right)^{2/n} \, \frac{1}{M_*}, 
\end{equation}
where $M_* \sim$ 1 TeV is the fundamental scale of gravity, 
$M_{\rm pl} = 10^{18}$ GeV, and $n$ is the number of extra dimensions. With 
this 
factorizable geometry the case of one extra dimension is clearly excluded 
since gravity would then be modified  at the scale of our solar system. 
However, 
for $n\geq 2$, $r$ is sufficiently small  (the fundamental 
Planck scale is lowered all the way to the TeV scale) and the model 
is not excluded by short distance gravitational measurements.
A more compelling 
scenario requires curvature to spill into the extra dimension \cite{lisa}. 
Within this framework the background metric is not flat along the 
extra coordinate, rather it is a slice of anti de Sitter space, due to a 
negative bulk cosmological constant balanced by the tension of two branes. 
In this non-factorizable geometry, the curved nature of the spacetime causes 
the physical scale on the two branes to be different, and exponentially 
suppressed in the negative tension brane. Such exponential suppression can 
then naturally explain why the physical scales observed are so much smaller 
than the Planck scale. Variants of this solution have been discussed by 
many authors \cite{+rs}. These models 
make dramatic predictions which can be 
directly confronted by current and future collider experiments \cite{pheno}, 
as well as cosmological observations \cite{cosmological}. The search for 
extra-dimension footprints in 
collider data has already started. However, as yet no 
observational evidence has been found  \cite{rs-exp}. 

Another seemingly different, but perhaps closely related 
subject is the apparent lack of a high energy cutoff in the cosmic ray (CR) 
spectrum. Over the last few years, several giant air showers have been 
detected \cite{yd}, with no sign of the expected 
Greisen-Zatsepin-Kuz'min (GZK) cutoff \cite{gzk}.
Initiated by single high energy particles hitting the atmosphere, these are 
large pancake-shaped slabs of high energy particles which hit the ground 
at nearly the speed of light and can cover areas of many square kilometers.
The origin and nature of the progenitors is, at present, a deep  
mystery \cite{es}. Protons with energies above the GZK cutoff lose energy 
rapidly via inelastic collisions with the cosmic microwave background (CMB) 
and thus presumably must come from a nearby source. This seems 
unlikely \cite{es}. A 
typical nucleus of the cosmic radiation is subject to photodisintegration 
from blue-shifted microwave photons, losing about 3-4 nucleons per 
traveled Mpc \cite{nuclei}. Gamma rays of the appropriate energy have a short 
mean 
free path for 
creating electron-positron pairs \cite{gamma}. 
Although neutrinos can propagate through the CMB essentially uninhibited, at these energies the atmosphere is still transparent, and most of them interact
in the Earth if at all.
The difficulties encountered in 
identifying a known particle as candidate have motivated suggestions in favor 
of ``exotic'' massive neutral hadrons, whose range is not limited by 
interactions with the CMB \cite{CFK}. However,  the latter predicts a correlation between 
primary arrival directions and the high redshift sources, which is not 
supported by the data set now available \cite{sigletal}.  
On a different track, it was recently put forward that extra dimensions may 
in principle hold the key to overcome this puzzle \cite{neutrinos}. In this 
article we shall explore this fascinating possibility.

It has long been known that any early universe phase transition occurring 
after inflation (say with symmetry breaking temperature $T_c$), which leaves 
unbroken a $U(1)$ symmetry group, may produce magnetic monopoles \cite{mono}.
For instance, minimal $SU(5)$ breaking may lead to ``baryonic monopoles'' 
of mass $M \sim T_c/ \alpha$, with magnetic charge $U(1)_{\rm EM}$ and 
chromomagnetic (or color-magnetic charge) $SU(3)_{\rm C}$ \cite{g}. Here 
$\alpha$ stands for the fine structure constant at scale 
$T_c$. These monopoles easily pick up energy from the magnetic fields 
permeating the universe and can traverse unscathed through the primeval 
radiation. Thus, they are likely to generate extensive 
air showers  \cite{tom}.\footnote{The idea of monopoles as 
constitutents of primary cosmic radiation is 
actually quite old, it  can be traced back at least as far as 
1960 \cite{porter}.} 
Before proceeding further, it is important to point out 
that if the monopoles are formed at the usual 
grand unification (GUT) scale  $\sim 10^{15}$ GeV, the energy density 
overcloses the universe.
Thus, to avoid this effect the symmetry breaking scale associated with the 
production of monopoles has to be shifted to lower energies. Remarkably, 
if the GUT scale is at $\sim 10^9$ GeV, one would end up with 
an abundance of relativistic monopoles well below the closure limit, and yet 
potentially measurable to explain the tail of the CR-spectrum. In addition, 
for such a critical temperature the observed flux of ultra high energy CRs 
is below the flux allowed by the Parker 
limit \cite{parker}. Moreover, the CR flux does not violate 
the upper bound for the monopole flux based on preliminary results quoted by 
the AMANDA Collaboration 
(see Fig. 1) \cite{amanda}. Unfortunately, contrary to the observed CR 
arrival directions, the expected 
flux of relativistic monopoles is highly anisotropic, pointing towards the 
magnetic lines near the Earth \cite{escobar}.

\begin{figure}
\label{0}
\begin{center}
\epsfig{file=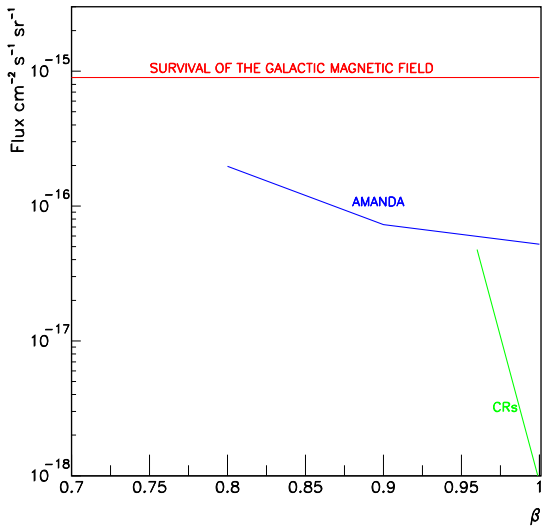,width=8.cm,clip=} 
\caption{The upper flux limit of monopoles (as a function of the 
relativistic parameter $\beta$) as observed by the AMANDA experiment,   
is shown together with the Parker bound and the 
CR flux assuming that the primaries at the end of the spectrum 
($E \geq 5 \times 10^{19}$ eV) have masses $\sim 10^{10}$ GeV.}
\end{center}
\end{figure}

In the multidimensional models, the low-scale unification enables the 
production of light-mass monopoles, say $M\sim100$ TeV. Furthermore, the 
physical embodiment of these theories allows a natural generalization of 
the 't Hooft-Polyakov monopole providing a convenient set of 
representations for D1-branes ending on  D3-branes, and consequently 
even lighter monopoles. 
Note, however,  that direct searches at 
accelerators pretty much exclude masses below a few hundreds of GeV, whereas 
bounds stemming from quantum effects on current observables turn out to be 
$\sim 1$ TeV \cite{alvaro}. 
The light-mass monopoles could lose and gain energy as 
they random-walk towards the Earth. The maximum energy attainable before 
hitting the atmosphere is roughly $10^{25}$ eV \cite{wkwb}. Therefore,  
these ``particles''  would be ultra-relativistic, and the expected flux 
has no imprint of correlation with the local magnetic field. 

To mimic a shower initiated by a proton the monopole must transfer nearly
all of its energy to the atmospheric cascade in a very small distance.
The large inertia of a massive monopole makes this impossible
if the cross-section is typically strong, $\sim$ 100 mb.
Wick, Kephart, Weiler and Biermann (WKWB) \cite{wkwb} have recently pointed 
out that this problem can be avoided in models in which the baryonic
monopole consists of $q$-monopoles confined by strings of
chromomagnetic flux. To describe the interactions of such
a monopole in air, WKWB have developed a model based on the four following 
axioms: i) before hitting the atmosphere the monopole-nucleus cross 
section is roughly hadronic $\sigma_0 \sim \Lambda_{\rm QCD}^{-2}$ 
(unstretched state), attaining a geometric growth after the impact; ii) in 
each interaction an ${\cal O} (1)$ fraction of the exchanged energy goes into 
stretching the  chromomagnetic strings of the monopole; iii) the 
chromomagnetic strings (of tension $T\sim \Lambda_{\rm QCD}^{-1}$) can only 
be broken to create monopole-antimonopole pairs (a process 
highly supressed and consequently ignored); iv) the average fraction of energy 
transferred to the shower in each interaction is soft 
$\Delta E/E \equiv \eta  \approx \Lambda_{\rm QCD}/M$.  

Generally speaking, in this set up the monopole will penetrate deeply into the 
atmosphere (the cross section is comparable to that of a high energy proton). 
However, since the geometrical cross-section grows 
proportionally with the Lorentz factor $\gamma$, the interaction length (after 
the impact) shrinks to a small fraction of the depth of the first 
interaction. Stated 
mathematically, the unstretched monopole's string length, 
$L \sim \Lambda^{-1}$, increases by $\delta L = \Delta E/T$. Recalling that 
nearly all  of the exchanged energy goes into stretching  the color magnetic strings, the fractional increase in the length is $\delta L/L = \gamma$, yielding $\sigma_1 \sim (1+\gamma)/\Lambda_{\rm QCD}^2$. Now, 
 the total mean free path after the $N$-th interaction reads,  
\begin{equation}
\lambda_N \sim \frac{1}{\sigma_N \,n_{\rm nuc}} \sim \frac{\Lambda^2_{\rm QCD}}
{(1 + \sum_{j=1}^N \gamma_j) \,n_{\rm nuc}} \sim \frac{\Lambda^2_{\rm QCD}}
{N\,\gamma \,n_{\rm nuc}},
\end{equation}
where we have 
assumed a constant density of nucleons $n_{\rm nucl} \approx (4/3)\, \pi\, A \,R_0^3$
and we have used the approximation 
$\gamma_N \sim (1 - \Lambda_{\rm QCD}/M)^N 
\gamma \sim \gamma$. Here $A$ stands for the mass number of an atmospheric 
nucleus, and $R_0 \approx 1.2 - 1.5$ fm. It should also be stressed 
that for $N = \eta^{-1}$ the approximation has an error  
bounded by $\lim_{N \rightarrow \infty} (1- N^{-1})^N = e^{-1}$. 
For $\eta^{-1} \gg 1$, the total 
energy traveled between the first interaction and 
the $\eta^{-1}$-th interaction is then
\begin{equation}
\Delta X \sim \frac{\Lambda_{\rm QCD}^2}{\gamma \,n_{\rm nuc}} 
\sum_{N=1}^{\eta^{-1}} \frac{1}{N} \sim \frac{\Lambda_{\rm QCD}^2}{\gamma \,n_{\rm nuc}} \ln \eta^{-1}.
\end{equation}
Note that the mean free path for all secondary interactions is 
${\cal O} (1/\gamma)$ compared to the first one. All in all, a baryonic 
monopole encountering the atmosphere will diffuse like a proton, producing a 
composite heavy-particle-like cascade after the first interaction.

\begin{figure}
\label{1}
\begin{center}
\epsfig{file=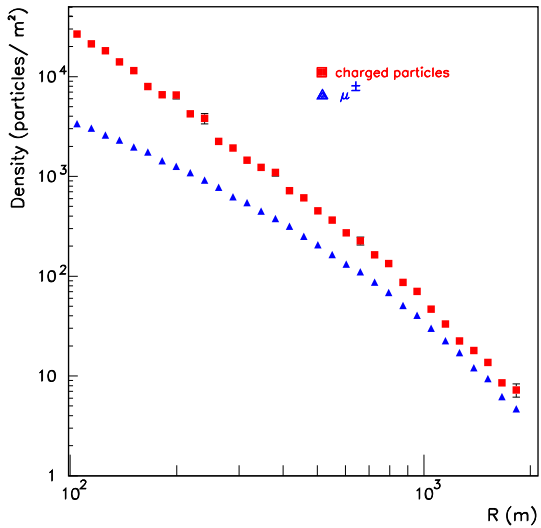,width=8.cm,clip=} 
\caption{ Lateral distributions of charged particles and muons
from {\sc aires} simulations of a $100$ EeV monopole with $M = 100$ TeV 
as a function of the distance to the shower core $R$.
The error bars (obscured by the points themselves in most cases) indicate 
the RMS fluctuations of the means.}
\end{center}
\end{figure}

To examine the signature of such a cascade, we carried out a Monte Carlo simulation of monopole showers 
{\it \`a la} WKWB using the {\sc aires} program 
(version 2.2.1) \cite{sergio}. Specifically, several sets of 
proton ``clumps'', 
each containing $M/\Lambda_{\rm QCD}$, 
were injected at 100 km a.s.l with the first interaction point
fixed according to the proton mean free path.
The sample was distributed in the energy
range of $1 \times 10^{18}$ eV up to $3 \times 10^{20}$ eV, and was equally
spread in the interval of 0$^{\circ}$ to 60$^{\circ}$ zenith angle at
the top of the atmosphere. All shower particles with energies above the 
following thresholds were
tracked: 750 keV for gammas, 900 keV for electrons and positrons, 10
MeV for muons, 60 MeV for mesons and 120 MeV for nucleons.
The hadronic interaction was modelled with 
the {\sc sibyll} package \cite{sibyll}. 
The results of these simulations were processed with the help of the 
{\sc aires} analysis programs.

\begin{figure}
\label{2}
\begin{center}
\epsfig{file=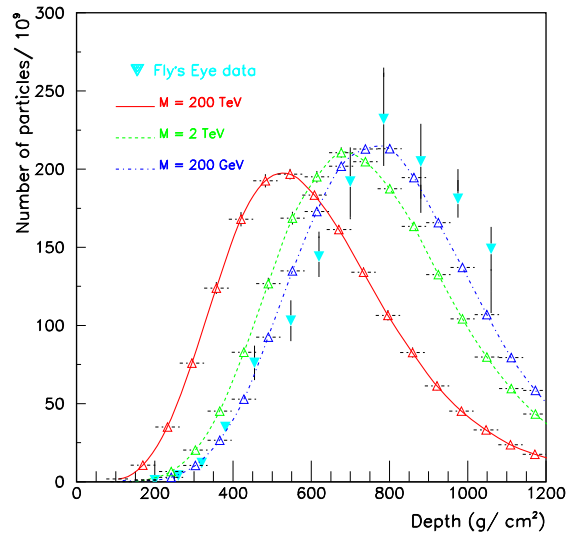,width=8.cm,clip=} 
\caption{Atmospheric cascade development of 300 EeV monopole induced 
showers, superimposed over the Fly's Eye data. The error bars in the 
simulated curves indicate the RMS fluctations of the means.}
\end{center}
\end{figure}

The resulting lateral distributions from a vertically incident 
monopole 
of 100 EeV ($\gamma \equiv 10^{6}$) for muons and charged 
particles are presented in Fig. 2. A distinctive signature of this kind 
of shower is the great number of muons among all charged particles. 
This feature was observed in one not well understood ``super-GZK'' event 
\cite{yakutsk}. Roughly speaking, a magnetic monopole could then be a 
candidate primary for the highest energy Yakutsk event.
However, WKWB-monopoles associated with a symmetry breaking at $T_c\sim 1$ 
TeV certainly cannot explain all features of the  data at the end 
of the spectrum. This is illustrated in Figs. 3 and 4. In Fig. 3 we show the 
longitudinal development of monopole showers superimposed over the 
experimental data of the world's highest energy cosmic ray to date \cite{FE}.
To get some numerical estimates we analyzed the data by means of a $\chi^2$ 
test \cite{pdg}. We assume that the set of measured values by Fly's Eye are 
uncorrelated (any depth measurement is independent of any other), and 
make use of the quantity 
\begin{equation}
\chi^2 \equiv \sum_{j=1}^q \frac{|x_j - \alpha_j|^2}{\sigma_{x_j}^2},
\end{equation}
where $q$ is the total number of points in the analysis, 
$\sigma_{x_j}$ is the error on the $x_j$th coordinate, $x_j$ is the measured 
value of the coordinate, and $\alpha_j$ the (hypothetical) true value of the 
coordinate. For masses of a few hundred GeV, the obtained $\chi^2$ increases 
with rising mass from 13.9 to 58.4. Our analysis indicates that masses above 
600 GeV are excluded at more than 99 \% C.L. On the other hand, WKWB 
monopoles of masses around 200 GeV become an alternative explanation for the 
Fly's Eye event. It is important to stress that a monopole mass 
$\approx 200$ GeV is not favored by D\O $\,\,$ data \cite{k}, although one 
should keep in mind that these bounds are quite model dependent. Moreover, in 
view of the wide variety of uncertainties in the Fly's Eye event (the total 
error in the energy determination is 93 EeV \cite{FE}), one may be 
excused for reserving final judgment until more data is available.   

A better understanding of the present situation 
needs the analysis of the evolution of the shower maximum $X_{\rm max}$ with 
energy. To this end, the charge multiplicity (essentially electrons 
and positrons)  was used to determine the number of particles and the 
location of $X_{\rm max}$ by means of four parameter fits to the 
Gaisser-Hillas function \cite{sergio}. The situation is summarized by 
displaying the mean  $X_{\rm max}$ as a function of the logarithm of the 
primary energy in Fig. 4. 
It is clear that despite its deep penetration, the monopole cascade develops 
much faster than a proton shower \cite{prdhi}. 
It can be seen by inspection that the $X_{\rm max}$ values produced by 
ultra-relativistic monopoles ($E \agt 10^{19}$ eV) with masses $\agt 200$ TeV
are inconsistent with those reported by the 
Fly's Eye experiment, whereas the $X_{\rm max}$ values of showers induced by 
lighter monopoles, $M \alt 500$ GeV, are within 1 standard deviation of the 
scarce ``super-GZK'' data \cite{yd}.

\begin{figure}
\label{3}
\begin{center}
\epsfig{file=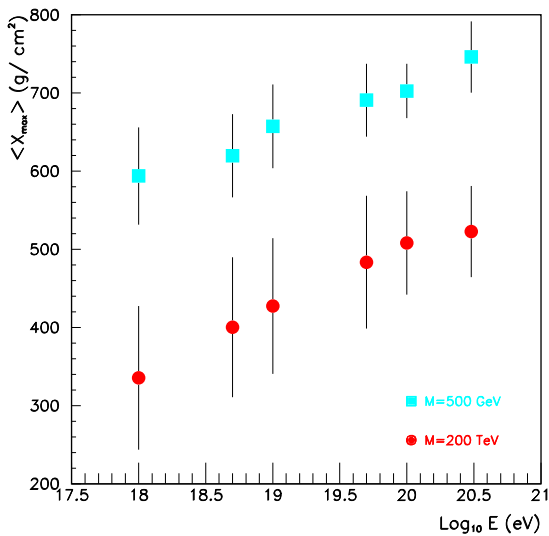,width=8.cm,clip=} 
\caption{Average slant depth of maximum of showers initiated by monopoles. 
The error bars indicate the RMS fluctuations of the means.}
\end{center}
\end{figure}

Whether or not the laws of physics should be formulated in
more than four dimensions is still unclear. 
The possible existence of compact large extra dimensions brings with it  
low energy phase transitions \cite{gut}, providing a profitable arena for 
baryonic light--mass monopole production. If this is the case, the 
monopoles could be accelerated to ultra-relativistic energies as they 
roam through space, inducing extensive air showers from time to time 
after hitting  Earth atmosphere. As we have discussed in this 
article, the atmospheric cascade of monopoles with $M \alt 500$ GeV could 
reproduce quite well the main features of the recorded giant cosmic ray 
showers. Certainly, more data is needed to test the WKWB hypothesis. 
Forthcoming ground arrays and satellites, such as the Auger 
Observatory \cite{auger}, the next SCROD \cite{scrod}, and 
EUSO/OWL/AirWatch \cite{euso,owl,airwatch}, 
will help to increase the CR sample and more precise limits on the air shower 
observables will be available, shedding light on the ideas discussed in this 
paper.

\acknowledgements{
We would like to thank Haim Goldberg, Carlos Nu\~nez, and Tom Weiler for 
useful discussions. We are also indebted to Gia Dvali for insightful 
comments on the manuscript, and to Peter Nie{\ss}en, Doug Cowen, 
and Christian Spiering for pointing us to the unpublished 
AMANDA results. The work was supported by CONICET (Argentina) and the 
National Science Foundation.}


\begin{thebibliography}{99}


\bibitem{renacimiento} Variations of Kaluza--Klein 
theory that motivated the renaissance of higher dimensional 
models are discussed in, K. Akama, Lect. Notes Phys. {\bf 176}, 267 (1982) 
[hep-th/0001113]; V. Rubakov and M. Shaposhnikov, 
Phys. Lett. {\bf 125B}, 136
(1983); M. Visser, Phys. Lett. {\bf B159}, 22 (1985);
M. Pavsic, Class. 
Quant. Grav. {\bf 2}, 869 (1985); Phys. Lett. A {\bf 107}, 66 (1985);   
G. W. Gibbons and D. L. Wiltshire, Nucl. Phys. B {\bf 287}, 717 (1987);
I. Antoniadis, Phys. Lett. B {\bf 246} (1990) 377. See, A. 
P\'erez-Lorenzana, [hep-ph/0008333], for a recent survey 
and bibliography on the subject.
  



\bibitem{trapping} 
G. Dvali and M. Shifman, Phys. Lett. B {\bf 396}, 64 (1997), 
{\it erratum ibid} {\bf 407}, 452 (1997);
M. Gogberashvili, Mod. Phys. Lett. A {\bf 14}, 2025 (1999);
L. Randall and R. Sundrum, Phys. Rev. Lett. {\bf 83}, 
4690 (1999);  M. Gogberashvili, [hep-ph/9908347]; 
B. Bajc, G. Gabadadze, Phys. Lett. B {\bf 474}, 282 (2000);
C. Grojean,
Phys. Lett. B {\bf 479}, 273 (2000); 
G. Dvali, G. Gabadadze, [hep-th/0008054].

\bibitem{nima} N. Arkani-Hamed, S. Dimopoulos and G. Dvali, Phys.
Lett. B {\bf 429}, 263 (1998); I. Antoniadis, N. Arkani-Hamed,
S. Dimopoulos and G. Dvali, Phys. Lett. B {\bf 436}, 257 (1998).

\bibitem{lisa}  L. Randall and R. Sundrum, Phys. Rev. 
Lett. {\bf 83}, 3370 (1999).

\bibitem{+rs} The number of papers discussing variants of the 
Randall--Sundrum scenario is already very large. Some key papers are: 
 J. Lykken and L. Randall, JHEP {\bf 0006}, 014 (2000)
[hep-th/9908076];   
N. Kaloper,  Phys. Rev. D {\bf 60}, 123506 (1999)
[hep-th/9905210]; C. Cs\'aki, M. Graesser, C. Kolda, J. Terning,
Phys. Lett. B {\bf 462}, 34 (1999) [hep-ph/9906513];
J. M. Cline, C. Grojean and G. Servant, Phys. Rev. Lett. {\bf 83},
4245 (1999) [hep-ph/9906523]; 
N. Arkani-Hamed, S. Dimopoulos, G. Dvali and N. Kaloper, Phys. Rev. Lett.
{\bf 84}, 586 (2000) [hep-th/9907209];
W. D. Goldberger and M. B. Wise, Phys. Rev. 
Lett. {\bf 83}, 4922 (1999) [hep-ph/9907447]; 
P. Kraus,  JHEP {\bf 9912} 011 (1999)
[hep-th/9910149];
J. Garriga and T. Tanaka, Phys. Rev. Lett.
{\bf 84} 2778 (2000) [hep-th/9911055];
N. Arkani-Hamed, S. Dimopoulos, G. Dvali, and N. Kaloper, [hep-ph/9911386];
C. Cs\'aki, M. Graesser, L. Randall and 
J. Terning, Phys. Rev. D {\bf 62}, 
045015 (2000) [hep-ph/9911406]; 
J. Garriga and M. Sasaki, Phys. Rev. D {\bf 62}, 
043523 (2000) [hep-th/9912118];
K. Koyama, J. Soda, Phys. Lett. B {\bf 483}, 432 (2000) [gr-qc/0001033];
S. B. Giddings, E. Katz and L.
Randall, JHEP {\bf 0003}, 023 (2000) [hep-th/0002091];
S. Nojiri, S. D. Odintsov and S. Zerbini, Phys. Rev. D
{\bf 62}, 064006 (2000) [hep-th/0001192]; S. W. Hawking, T. Hertog and 
H. S. Reall, Phys. Rev. D {\bf 62}, 043501 (2000) [hep-th/0003052]; 
C. Barcel\'o and M. Visser,
Phys. Lett. B {\bf 482}, 183 (2000) [hep-th/0004056];
L. Anchordoqui, C. Nu\~nez and K. Olsen, JHEP (to be published) 
[hep-th/0007064]; C. Cs\'aki, M. L. 
Graesser and G. D. Kribs [hep-th/0008151].

\bibitem{pheno} G. F. Giudice, R. Rattazzi 
and J. D. Wells, Nucl. Phys. B {\bf
544}, 3 (1999) [hep-ph/9811291]; S. Cullen, M. Perelstein and 
M. E. Peskin, [hep-ph/0001166].

\bibitem{cosmological} S. Cullen and M. Perelstein, Phys. Rev. Lett. {\bf 83},
268 (1999) [9903422]; V. Barger, T. Han, C. Kao, and R.-J. Zhang, 
Phys. Lett. B {\bf 461}, 34 (1999) [hep-ph/9905474]; 
N. Kaloper, and A. R. Liddle, Phys. Rev. D {\bf 61}, 123513 (2000) 
[hep-ph/9910499]; R. Maartens, D. Wands, B. Bassett, I. Heard, Phys. Rev. D 
{\bf 62}, 041301 (2000) [hep-ph/9912464]; N. Arkani-Hamed, S. Dimopoulos, 
N. Kaloper, and R. Sundrum, Phys. Lett. B {\bf 480}, 193 (2000) 
[hep-th/0001197];
S. Cassisi, V. Castellani, S. Degl'Innocenti, G. Fiorentini and B. Ricci, 
[astro-ph/0002182]; 
Z. K. 
Silagadze, [hep-ph/0002255]; 
R. Maartens, Phys. Rev. D {\bf 62}, 084023 (2000) [hep-th/0004166];
D. Langlois, R. Maartens and D. Wands, Phys. 
Lett. B (in press) [hep-th/0006007]; A. Mazumdar, [hep-ph/0007269]; 
[hep-ph/0008087]; C. Gordon, R. Maartens, [hep-th/0009010]; 
S. Kobayashi, K. Koyama and J. Soda, [hep-th/0009160].


\bibitem{rs-exp}
M. Acciarri et al. (L3 Collaboration), Phys. Lett. B {\bf 470},
281 (1999);
C. Adloff et al. (H1 Collaboration),
[hep-exp/0003002];
B. Abbott et al. (D\O $\,\,$ Collaboration), [hep-ex/0008065].

\bibitem{yd} S. Yoshida and H. Dai, J. Phys. G {\bf 24}, 905 (1998).

\bibitem{gzk} If the cosmic ray sources are all at cosmological distances, 
then the observed energy spectrum should not extend, except at greatly 
reduced intensity, beyond $10^{20}$ eV. K. Greisen, Phys. Rev. Lett. 
{\bf 16}, 748 (1966); G. T. Zatsepin, and V. A. Kuz'min, Pis'ma Zh. \'Eksp. 
Teor. Fiz. {\bf 4}, 114 (1966) [JETP Lett. {\bf 4}, 78 (1966)].

\bibitem{BS} For a thorough and up-to-date survey on the origin
of the highest energy CRs the reader is referred 
to, P. Bhattacharjee, and G. Sigl, Phys. Rep. {\bf 327}, 109 (2000).

\bibitem{es} J. W. Elbert and P. Sommers, Astrophys. J. 
{\bf 441} (1995) 151.


\bibitem{nuclei} L. N. Epele and E. Roulet, J.
High Energy Phys. {\bf 10}, 009 (1998); F. W. Stecker, and M. H.
Salamon, Astrophys. J. {\bf 512} 521 (1999).

\bibitem{gamma} R. J. Protheroe and P. L. Biermann, Astropart. Phys. {\bf 6} 
45 (1995),  {\it erratum ibid.} {\bf 7}, 181 (1997).  


\bibitem{CFK} D. J. H. Chung, G. R. Farrar and E. W. Kolb, Phys. Rev.
D {\bf 59}, 015021 (1999).


\bibitem{sigletal} G. Sigl, D. F. Torres, L. A. Anchordoqui and G. E. 
Romero, [astro-ph/0008363].

\bibitem{neutrinos}  The virtual exchange of bulk gravitons 
(Kaluza--Klein modes) yields extra contributions to the neutrino-nucleon 
cross section, producing earlier development of a neutrino induced shower. 
G. Domokos and S. Kovesi-Domokos, Phys. Rev. 
Lett. {\bf 82}, 1366 (1999);
S. Nussinov and R. Shrock, Phys. Rev. D {\bf 59} 105002 
(1999); P. Jain, et al., 
Phys. Lett. B {\bf 484}, 267 (2000); C. Taylor, A. Olinto and 
G. Sigl, [hep-ph/0002257]; G. Domokos, S. Kovesi-Domokos and P. T. 
Mikulski [hep-ph/0006328]. The model, however, is not free of problems. See, 
M. Kachelrie{\ss} and M. Plumacher, [astro-ph/0005309].


\bibitem{mono} G. 't Hooft, Nucl. Phys. B {\bf 79}, 276 (1974); 
A. M. Polyakov, JETP Lett. {\bf 20}, 194 (1974).


\bibitem{g} A. S. Goldhaber, Phys. Rep. {\bf 315}, 83 (1999).

\bibitem{tom}T. W. Kephart and T. J. Weiler, Astropart. Phys. 
{\bf 4}, 271 (1996); 
Nucl. Phys. (Proc. Suppl.) {\bf 51B}, 218 (1996).

\bibitem{porter} N. A. Porter, Nuovo Cim. {\bf 16}, 958 (1960). 


\bibitem{parker} This bound requires that there  be not so many monopoles 
around as to effectively ``short out'' the galactic magnetic field.
M. S. Turner, E. N. Parker
and T. Bogdan, Phys. Rev D {\bf 26}, 1296 (1982).


\bibitem{amanda} P. Nie{\ss}en, private communication. 


\bibitem{escobar} C. O. Escobar and R. A. V\'azquez, Astropart. Phys. {\bf 10},
197 (1999).


\bibitem{alvaro} A. De R\'ujula, Nucl. Phys. B {\bf 435}, 257 (1995). 
The value for the lower limit of the mass of the monopole is still under 
debate. See for instance, L. Gamberg, G. R. Kalbfleisch and K. A. 
Milton, [hep-ph/9906526].  


\bibitem{wkwb} S. D. Wick, T. W. Kephart, T. J. Weiler 
and P. L. Biermann, Astropart. Phys. (to be published) [astro-ph/0001233].


\bibitem{sergio} S. J. Sciutto, in {\it
Proc. XXVI International Cosmic Ray Conference}, (Edts. D. Kieda, M. Salamon,
and B. Dingus, Salt Lake City, Utah, 1999) vol.1, p.411, [astro-ph/9905185]. 


\bibitem{sibyll} R. S. Fletcher, et al., Phys. Rev. D {\bf 50}, 5710 (1994).


\bibitem{yakutsk}  N. N. Efimov, et al., in
Astrophysical Aspects of the Most Energetic Cosmic Rays, (eds. M.
Nagano, F. Takahara, World Scientific 1991), p.434. See also,
E. E. Antonov et al.,  Pis'ma Zh. \'Eksp. Teor. Fiz.
{\bf 69}, 614 (1999) [JETP Lett. {\bf 69}, 650 (1999)].

\bibitem{FE} D. J. Bird et al., Astrophys. J. {\bf 441}, 144 (1995).


\bibitem{pdg} For a concise review see the probability and statistics 
sections of, R. M. Barnett et al. (Particle Data Group), 
Phys. Rev. D {\bf 54}, 155 (1996).  


\bibitem{k} G. R. Kalbfleisch et al., [hep-ex/0005005].


\bibitem{prdhi} L. A. Anchordoqui, M. T. Dova, L. N. Epele and S. J. Sciutto, 
Phys. Rev. D {\bf 59}, 094003 (1999).


\bibitem{gut} K. R. Dienes, E. Dudas and T. Gherghetta, Phys. Lett. B 
{\bf 436}, 55 (1998).

\bibitem{auger} {\tt http://www.auger.org/admin/}

\bibitem{scrod} {\tt http://hepnt.physics.neu.edu/scrod/}

\bibitem{euso}{\tt http://ifcai.pa.cnr.it/ifcai/euso.html}

\bibitem{owl} {\tt http://owl.gsfc.nasa.gov}

\bibitem{airwatch} {\tt http://www.ifcai.pa.cnr.it/\lower3pt\hbox{\string~}AirWatch/}

\end{thebibliography}
\end{document}